\documentstyle[aps,epsfig,twocolumn]{revtex}
\begin{document}
\draft
\wideabs{
\title{Control of Coherent Acoustic Phonons}

\author{\"{U}mit \"{O}zg\"{u}r, Chang-Won Lee, and Henry O. Everitt$^{\rm *}$
\footnotetext{$^{\rm *}$Also U.\ S.\ Army Research Office,
 Research Triangle Park, NC 27709-2211 Electronic mail:
everitt@arl.aro.army.mil} 
}
\address{Department of Physics, Duke University, Durham, NC 27708}
\date{\today}
\maketitle
\begin{abstract}
Using sub-picosecond optical pump-probe techniques, coherent zone-folded
longitudinal acoustic phonons (ZFLAPs) were generated and controlled in an
InGaN multiple quantum well  structure. A one-pump, one-probe differential
transmission technique revealed that carriers injected near the barrier
band edge were quickly captured into the quantum wells and generated strong
coherent ZFLAP oscillations. Two-pump differential transmission was used to generate and 
control coherent ZFLAP oscillations through the relative timing and amplitude of the 
two pump pulses. Enhancement and suppression of ZFLAP oscillations were demonstrated, including
complete cancellation of generated acoustic phonons for the first time in 
any material system. Coherent control was used to demonstrate that ZFLAPs are generated differently 
in InGaN multiple quantum wells than in GaAs/AlAs superlattices.
\end{abstract}

}

The techniques of ultrafast optical spectroscopy have provided unprecedented
capabilities to generate and control coherent quantum mechanical processes 
and to examine fundamental physical phenomena such as relaxation, dephasing, 
and squeezing. The ability to control coherent behavior is well established 
in atomic and molecular systems \cite{WarrenSCI93,GaetaPRL94,WeinachtNAT99}. 
Demonstrations of control of coherent behavior in condensed matter systems 
has been more problematical \cite{GrossPRB94}, primarily because of
the much faster dephasing times and difficulty of manipulating coherent 
states on time scales much shorter than this. Recent demonstrations of 
control and entanglement of excitonic and biexcitonic states in 
semiconductor quantum dots \cite{BonadeoSCI98} suggest that such problems 
are not insurmountable, and that control of coherent phenomena in condensed 
matter systems is achievable.

Inevitably, electron-phonon interactions limit the dephasing time of electronic or
excitonic coherent states, so understanding and controlling the effects of phonons is
fundamental interest \cite{SchwabNAT00,BartelsAPL98}. Particularly challenging is
the optical investigation of acoustic
phonons in bulk semiconductors because the phonon dispersion relation permits only low frequency 
Brillouin scattering. However, it is well understood that a semiconductor
multiple quantum well (MQW) produces zone folding of the acoustic phonon branch so
that direct excitation is possible
\cite{BartelsAPL98,ColvardPRB85,KuznetsovPRL94,MizoguchiPRB99}. Recently, it has been
demonstrated that particularly strong coherent zone-folded longitudinal acoustic phonon (ZFLAP)   
oscillations can be generated and observed in InGaN MQW structures
\cite{SunAPL99,SunPRL00,OzgurAPL00}. In this paper, the optical mechanism for 
generating these coherent ZFLAP oscillations is shown to be impulsive, like the 
striking of a bell, much like the impulsive stimulated Raman scattering (ISRS) technique 
\cite{MerlinSSC97}. Demonstration of coherent ZFLAP control follows, including the first 
complete cancellation of generated acoustic phonons in InGaN or any other material 
system. The utility of coherent acoustic phonon control is exemplified by demonstrating that acoustic 
phonons are generated somewhat differently in 
InGaN MQWs than in GaAs/AlAs superlattices (SLs).

The MQW sample used here was grown by metalorganic chemical vapor deposition 
(MOCVD) at the University of California at Santa Barbara using a modified two-flow
horizontal reactor on double polished c-plane sapphire \cite{KellerJCG98}. It
consists of a $10$ 
period, $12$ nm per period MQW with $3.5$ nm wide In$_{0.15}$Ga$_{0.85}$N quantum
wells and $8.5$ nm wide In$_{0.05}$Ga$_{0.95}$N:Si barriers \cite{X-Ray}. The MQW
structure is
capped with a $100$ nm GaN layer and is grown on a $\sim 2 \mu$m GaN:Si  layer. The Si
doping concentration in the barriers is $\sim 10^{18}$ cm$^{-3}$.

In previous room
temperature measurements of carrier capture in this sample, photoluminescence (PL) and
photoluminescence excitation (PLE) spectroscopies identified the  energies of the
quantum well states and the barriers, respectively \cite{OzgurAPL00}. The peak
PL emission occurred at $2.99$ eV, corresponding to electron-hole recombination
from the lowest energy
quantum well subband states. The broad PLE absorption peak for emission at $2.99$ eV
occurred near $3.22$ eV, corresponding to electron-hole (e-h) pairs being generated
near the MQW barrier band edge.

\begin{figure}
\centerline{\resizebox{7cm}{!}{\hbox{\includegraphics{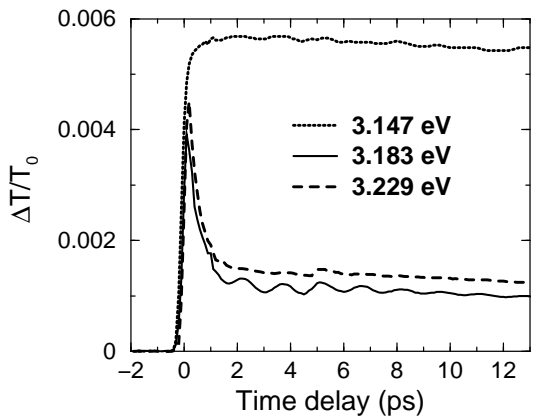}}}}
\caption{Single pump DT data for pump/probe wavelengths near ($3.229$ eV,
$3.183$ eV) and
below ($3.147$ eV) the barrier band edge. The feature at $5$ ps arises from a
pump pulse reflection
from the substrate.}
\label{Fig1}
\end{figure}

Wavelength-degenerate, sub-picosecond, pump-probe differential transmission (DT) has
been used to measure the electron capture time \cite{OzgurAPL00}. A strongly
wavelength-dependent
bi-exponential decay of the created carriers (Fig.~\ref{Fig1}) indicated that electrons were 
captured with a time constant between $0.31$ - $0.54$ ps. Electrons were most 
efficiently captured from 3D barrier states to 2D confined QW states when e-h pairs
were generated within $50$ meV of the $3.22$ eV barrier. Because of the greater 
density of states in the valence band, holes were captured too quickly to be 
measured. The strained material generates a large ($\sim 1$ MV/cm) piezoelectric
(PZE) field in the MQWs, \cite{TakeuchiAPL98} and the width of the PL signal and
region of efficient
carrier capture ($\sim 100$ meV) suggest sizable material inhomogenieties.

Remarkably strong damped oscillations in the DT signal are observed, suggesting the
generation of coherent phenomena in the quantum wells (Fig.~\ref{Fig1}). The damped 
oscillations always started at the peak of the pump pulse (t=$0$) and are described by
A$exp$(-t/$\tau$)$cos$($2\pi$t/P+$\pi$), whose phase term $\pi$ signifies that the
oscillations were always observed to start at a minimum. Observing similar phenomena
in similar MQW structures, Sun et al. \cite{SunAPL99,SunPRL00} concluded that 
such oscillations are the
manifestation of coherent ZFLAPs by demonstrating that the oscillation period P
increased linearly with increasing MQW period d as P = d/v$_s$, where v$_s$ is the
sound velocity. The ZFLAPs propagate along the c-axis of the wurtzitic InGaN. The
oscillations observed in our d = $12$ nm MQW sample occurred with a period P = $1.44$
ps (frequency = $f_0$ = $23$ cm$^{-1}$). Preliminary spontaneous Raman measurements 
confirmed 
a ZFLAP frequency of $23$
cm$^{-1}$, yielding a sound velocity of $8333$ m/s \cite{SoundVel}. The measured 
characteristic
damping time of these oscillations ($\tau$ = $12$ ps) sets a lower limit for the ZFLAP coherence time.

The sound velocity has not been measured directly in any InGaN material; however, the
value in bulk GaN along the c-axis is well established between $7990$ - $8020$ m/s 
\cite{DegerAPL98,YamaguchiJAP99}. Although InGaN is expected to be a softer
material with a slower sound
velocity, our low In content MQWs should not deviate much from the bulk GaN
values. Using the c-axis elastic stiffness tensor element (C$_{33}$) and the density
values for GaN ($395$ GPa, $6.087$ g/cm$^3$) and InN ($200$ GPa, $6.890$ 
g/cm$^3$), \cite{KimPRB96,WrightJAP97,Morkoc} the average sound velocity in the
MQW region can be estimated to be
$7888$ m/s. The difference in the measured and calculated values may arise from the
strain-induced piezoelectric field in the MQW region which tends to stiffen the
material and raise the sound velocity along the c-axis \cite{DegerAPL98}.

\begin{figure}
\centerline{\resizebox{7.5cm}{!}{\hbox{\includegraphics{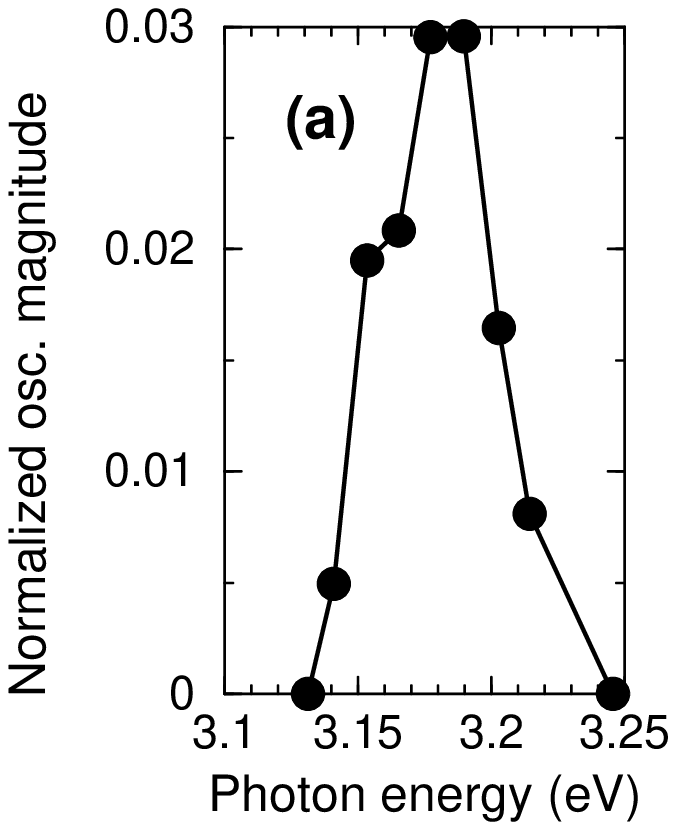}\includegraphics{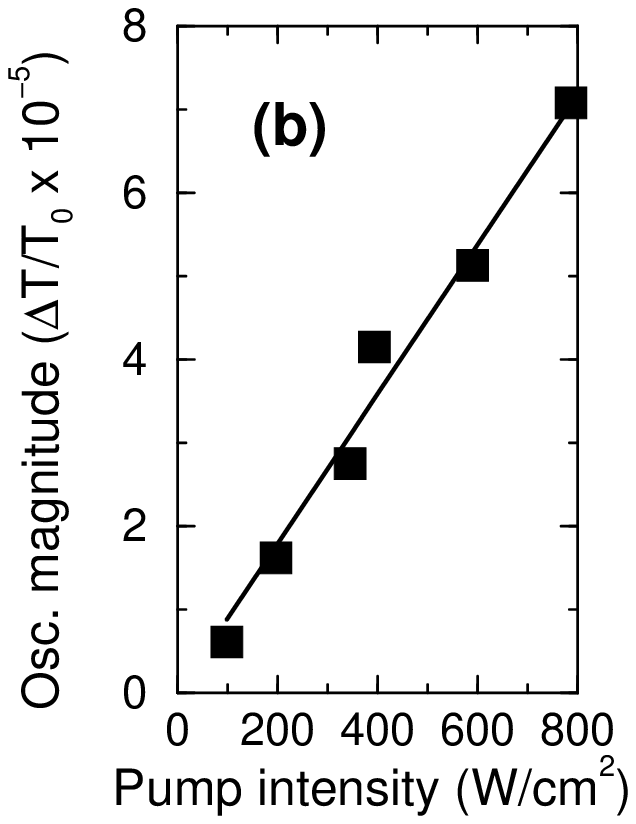}}}} 
\caption{(a) Strength of coherent ZFLAP oscillations as a function of
pump/probe energy, 
normalized with respect to the peak $\triangle$T/T$_0$ value at that energy. (b) Strength
of coherent ZFLAP
oscillations as a function of pump intensity.}
\label{Fig2}
\end{figure}

The coherent ZFLAP oscillations were strongest near the wavelength of most efficient
carrier capture ($3.18$ eV) and were only observed within $40$ meV of that energy 
[Fig.~\ref{Fig2}(a)]. Once carriers are captured into the wells, electrons and holes 
are separated by the strong PZE field and partially screen it. A nearly instantaneous change in the 
material stress results, impulsively inducing ZFLAP oscillations \cite{MerlinSSC97}. The screened PZE 
field also blueshifts the
absorption band edge through the quantum-confined Franz-Keldysh effect (QCFKE). The ZFLAP
oscillation modulates the carrier distribution that in turn modulates the strength of
the QCFKE and the absorption band shift, thereby producing oscillations in the DT 
signal. Thus, the span of pump/probe energies over which ZFLAP oscillations are
observed corresponds to the region of carrier capture and the width of the barrier 
absorption band edge.

To measure the dependence of the ZFLAP oscillation strength on
pump intensity, the $3.183$ eV pump and probe beams were focused to diameters of $75$
and $40$ microns, respectively. The time-averaged pump intensity was varied from $98$
to $787$ W/cm$^2$, while the probe intensity was kept at a constant $11$ 
W/cm$^2$. Fig.~\ref{Fig2}(b) demonstrates that the strength of the oscillations in this low 
excitation regime increases linearly with increasing pump intensity.

To achieve control of these coherent ZFLAPs, it is necessary to be able to manipulate
them on timescales shorter than the oscillation period. The pulse widths of the 
frequency-doubled, $80$ MHz Ti:Sapphire laser pulses ($ < 100$ fs) and the temporal 
resolution ($7$ fs) of the $1$ $\mu$m-per-step scanning motors of the delay lines for
the DT pump-probe apparatus are much smaller than the ZFLAP oscillation period
P. Coherent amplification and suppression of ZFLAPs may be 
demonstrated by the use of
two-pump, one-probe DT. The relative timing of the two pump pulses determines whether
the coherent ZFLAP oscillations add constructively (e.g. delay $\triangle$t = P) or
destructively (e.g. $\triangle$t = P/$2$). All three pulses were derived from the same
frequency doubled Ti:Sapphire laser pulse and were independently delayed with respect
to each other through the use of various beam splitters and delay stages. For
simplicity, the pump pulses were delayed relative to a fixed length probe pulse
pathway, and the two pump pulses were delayed relative to each other by the use of a
second delay stage mounted on the first. The three beams were made to overlap on the
sample by passing through a common lens in a coplanar geometry. The pump spot sizes 
were chosen to allow maximum pump intensity while ensuring the probe spot diameter was
well within those of the two pump spots.

\begin{figure}
\centerline{\resizebox{7cm}{!}{\hbox{\includegraphics{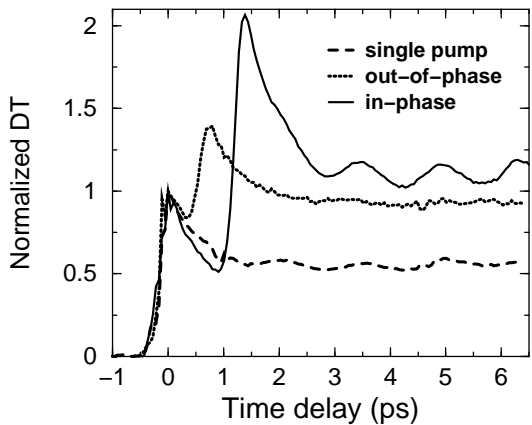}}}} 
\caption{Time-resolved one-pump and two-pump DT data. The two pumps were
one ZFLAP
oscillation period apart (in-phase, I$_2$/I$_1$ = $2.78$) or one-half ZFLAP
oscillation period apart (out-of-
phase, I$_2$/I$_1$ = $1.31$).}
\label{Fig3}
\end{figure}

\begin{figure}
\centerline{\resizebox{6.3cm}{!}{\hbox{\includegraphics{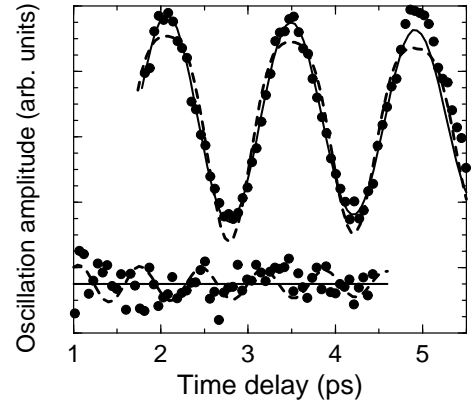}}}}
\caption{The data residuals of the fits for bi-exponential decay reveal the damped 
coherent ZFLAP oscillations for both two-pump cases in Fig.~\ref{Fig3}. The solid 
line represents the best fit assuming no overtone harmonics, and the 
dashed line illustrates the inferiority of the best fit including a second harmonic 
20\% 
as strong as the first.}
\label{Fig4}
\end{figure}

Fig.~\ref{Fig3} demonstrates the coherent amplification ("in-phase", 
$\triangle$t = P) and 
cancellation ("out-of-phase", $\triangle$t = P/$2$) of room temperature ZFLAP
oscillations. To examine the damped ZFLAP oscillations, a simple 
bi-exponential decay
was fit to the data, beginning at the peak of the second pump curve. The residual of
the fit reveals the damped oscillation of the ZFLAP, and a damped cosine oscillation
was fit to the residual as before to measure the amplitude A$_2$ and decay $\tau_2$ of
the oscillation (Fig.~\ref{Fig4}). The simple, damped oscillation following the 
second pump
suggests that the decay time of the second ZFLAP oscillation is the same as the first
($\tau_1$ = $\tau_2$ = $\tau$) and that phase coherence is maintained between the two
oscillations for at least $12$ ps.

\begin{figure}
\centerline{\resizebox{7cm}{!}{\hbox{\includegraphics{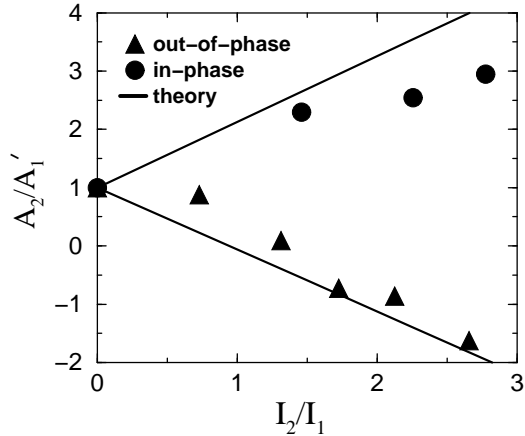}}}}
\caption{Strengths of the ZFLAP oscillations induced by the second pump,
relative to those induced
by the first pump, as a function of the ratio of the two
pump intensities.}
\label{Fig5}
\end{figure}

Even though the first ZFLAP oscillations have decayed somewhat by the time of the
second pump, it was observed that the intensity of the second pump must be 
\emph{increased}
relative to the first for exact amplitude doubling or cancellation. To provide a 
quantitative measure of the degree of amplification or suppression as a function of
relative pump intensities, the amplitude A$_2$, referenced to the amplitude of the
first ZFLAP oscillations at the time of the second pump [A$_1'$ = A$_1 exp$(-$\triangle$ 
t/$\tau$)], is plotted in Fig.~\ref{Fig5} for both the "in-phase" and 
"out-of-phase" cases. Negative values for the "out-of-phase" A$_2$/A$_1'$ indicate a
relative $\pi$ phase shift in amplitude, signifying that the second ZFLAP oscillations
are stronger than the first. Roughly the same pump intensity is required to double
the oscillations as to cancel them (I$_2$/I$_1$ $\approx$ $1.3$). Assuming the
strength of the oscillations increases linearly with pump intensity [Fig.~\ref{Fig2}(b)], the
predicted A$_2$/A$_1'$ = $1 \pm$ (I$_2$/I$_1$) $exp$($\triangle$ t/$\tau$) roughly agrees with
the data with no additional adjustable parameters for both in-phase (+) and 
out-of-phase (-) cases (Fig.~\ref{Fig5}). The slight discrepancy may be the result 
of PZE field
screening by the carriers captured after the first pulse, reducing the efficiency of
ZFLAP generation by the second pulse.

This ability to generate and control spectrally pure, very high frequency coherent ZFLAPs represents 
a new experimental regime for the phonon scientist. For the first time, a single acoustic phonon 
mode may be generated and controlled. Its spectral purity and strength derive from three attributes. 
First, impulsive DT does not generate acoustic phonons at 
frequencies given by the folded dispersion relation and the wavevector of the incident light 
\cite{BartelsPRL99}. In contrast with backscattering Raman these phonon doublets are of great 
spectroscopic benefit but represent a 
tremendous hinderance for phonon control. Second, previous studies of ZFLAP oscillations in 
GaAs/AlAs SLs using forward-scattering ISRS revealed a fundamental ZFLAP mode ($f$ = $f_0$) and its 
second harmonic ($f$ = 2$f_0$) \cite{BartelsAPL98}. Two-pump phonon control was able to suppress 
the fundamental ZFLAP mode while enhancing its second harmonic, even when the second harmonic was 
not apparent in the one-pump data. By contrast, InGaN MQW ZFLAP overtone harmonics were 
surprisingly missing in the one- and two-pump configurations. As suggested by Fig.~\ref{Fig4}, the 
two-pump data most strongly supports the hypothesis of no overtones, and statistical analysis 
rejects overtones of any amplitude, relative to the fundamental, of more than 5\%. Third, the 
one-pump, DT-induced ZFLAP oscillations were more than 10 times stronger than those reported for the 
GaAs/AlAs SL, even though the GaAs/AlAs SL had 4 times as many periods and was pumped almost 10 
times harder \cite{BartelsAPL98}. The likely explanation again lies with the intrinsic 
strain-induced PZE field in the InGaN MQWs. Photogenerated carriers partially screen the PZE field, 
suddenly relieving some of the incredible strain in the MQWs and impulsively generating strong, 
coherent ZFLAP oscillations. The relative strain relief is apparently much greater than the strain 
induced by the ISRS photogenerated carriers in unstrained GaAs/AlAs SLs.

The out-of-phase data in Fig.~\ref{Fig3} represents the first experimental demonstration of complete 
coherent acoustic phonon cancellation in any material system. The scientific and technological 
utility of single mode acoustic phonon generation and control is made even more attractive by the 
fact that it requires merely a standard 80 MHz, unamplified, mode-locked Ti:Sapphire laser, two 
delay stages, and a room temperature, short period MQW. Already, coherent control of ZFLAPs has 
revealed a fundamental difference in the generation of acoustic phonons in GaAs/AlAs SLs and InGaN 
MQWs. In addition, it is now possible to cancel coherent ZFLAPs one half period after their 
creation. The resulting single mode acoustic phonon impulse can propagate through the sample and 
may be traced temporally or mapped spatially. Conversely, repeated excitation with multiple in-phase 
pulses can continue to amplify the ZFLAP oscillations parametrically, permitting an investigation of 
phonon "gain", acoustic nonlinearity, and the role of the PZE field. These coherent techniques will 
enable unprecedented investigations of poorly understood acoustic phonon reflection and transmission 
at interfaces, with substrates, or in other materials to which the MQW is bonded. More generally, 
complex temporal pump pulse waveforms, created by genetic learning algorithms \cite{OmenettoJOSA99} 
and spatial light modulators\cite{WeinerRSI00}, can be used as a coherent terahertz ultrasonic 
transducer to create non-trivial phonon excitations to explore or control phonon mediated decay or 
dephasing of carriers and excitons.

The samples were grown by A. C. Abare, S. Keller, and S. P. DenBaars of the University of 
California, Santa Barbara. We thank R. Merlin for numerous valuable discussions, for suggesting the 
coherent control experiment with P. H. Bucksbaum, and for performing the coherent Raman measurements 
with J. Zhao. We thank A. C. Abare and M. J. Bergmann for the X-ray measurements and acknowledge 
additional helpful discussions with M. A. Stroscio and H. C. Casey, Jr. This
work was supported by ARO grant No. DAAH04-93-D-0002 and by DARPA/ARO grant
DAAH04-96-0076.


\begin{references}
\bibitem{WarrenSCI93}
W. S. Warren, H. Rabitz, and M. Dahleh, Science \textbf{259}, 1581 (1993).
\bibitem{GaetaPRL94}
Z. D. Gaeta, M. Noel, and C. R. Stroud, Jr., Phys. Rev. Lett. \textbf{73}, 636 (1994).
\bibitem{WeinachtNAT99}
T. C. Weinacht, J. Ahn, and P. H. Bucksbaum, Nature \textbf{397}, 233 (1999).
\bibitem{GrossPRB94}
P. Gross \emph{et al.}, Phys. Rev. B \textbf{49}, 11100 (1994).
\bibitem{BonadeoSCI98}
N. H. Bonadeo \emph{et al.}, Science \textbf{282}, 1473 (1998).
\textbf{282}, 1473 (1998).
\bibitem{SchwabNAT00}
K. Schwab, E. A. Henriksen, J. M. Worlock, M. L. Roukes, Nature \textbf{404}, 974 (2000).
\bibitem{BartelsAPL98}
A. Bartels, T. Dekorsy,  H. Kurz, and K. K\"{o}hler, Appl. Phys. Lett. \textbf{72}, 2844 
(1998).
\bibitem{ColvardPRB85}
C. Colvard \emph{et al.}, Phys. Rev. B \textbf{31}, 2080 (1985).
\bibitem{KuznetsovPRL94}
A. V. Kuznetsov and C. J. Stanton, Phys. Rev. Lett. \textbf{73}, 3243 (1994).
\bibitem{MizoguchiPRB99}
K. Mizoguchi, M. Hase, S. Nakashima, and M. Nakayama,  Phys. Rev. B \textbf{60}, 8262 
(1999).
\bibitem{SunAPL99}
C. K. Sun \emph{et al.}, Appl. Phys. Lett. \textbf{75}, 1249 (1999).
\bibitem{SunPRL00}
C. K. Sun, J. C. Liang, and X. Y. Yu, Phys. Rev. Lett. \textbf{84}, 179 (2000).
\bibitem{OzgurAPL00}
\"{U}. \"{O}zg\"{u}r \emph{et al.}, Appl. Phys. Lett. \textbf{77}, 109 (2000).
\bibitem{MerlinSSC97}
R. Merlin, Solid State Commun. \textbf{102}, 207 (1997).
\bibitem{KellerJCG98}
S. Keller \emph{et al.}, J. Cryst. Growth \textbf{195}, 258 (1998).
\bibitem{X-Ray}
X-ray measurements of the sample revealed the MQW period to be 12 nm, not the 8 nm
originally claimed by the growers and reported in [12].
\bibitem{TakeuchiAPL98}
T. Takeuchi \emph{et al.}, Appl. Phys. Lett. \textbf{73}, 1691 (1998).
\bibitem{SoundVel}
This value is significantly larger than the $6800$ m/s reported in 
\cite{SunAPL99,SunPRL00}, although the samples used in this study and in 
\cite{SunAPL99,SunPRL00} were nearly identical and grown by the same laboratory. 
Indeed, our results concurred with \cite{SunAPL99,SunPRL00} until we 
discovered the error in the reported MQW period \cite{X-Ray} which accounts for the 
discrepancy. 
\bibitem{DegerAPL98}
C. Deger \emph{et al.}, Appl. Phys. Lett. \textbf{72}, 2400 (1998).
\bibitem{YamaguchiJAP99}
M. Yamaguchi \emph{et al.}, J. Appl. Phys. \textbf{85}, 8502 (1999).
\bibitem{KimPRB96}
K. Kim, W. R. L. Lambrecht, and B. Segall, Phys. Rev. B \textbf{53}, 16310 (1996).
\bibitem{WrightJAP97}
A. F. Wright, J. Appl. Phys. \textbf{82}, 2833 (1997).
\bibitem{Morkoc}
H. Morko\c{c}, Nitride Semiconductors and Devices,  (Springer-Verlag, Berlin, Heidelberg,
New York, 1999).
\bibitem{BartelsPRL99}
A. Bartels, T. Dekorsy, H. Kurz, and K. K\"{o}hler, Phys. Rev. Lett. \textbf{82}, 1044 
(1999).
\bibitem{OmenettoJOSA99}
F. G. Omenetto, B. P. Luce, and A. J. Taylor, J. Opt. Soc. Am. B \textbf{16}, 2005 (1999).
\bibitem{WeinerRSI00}
A. M. Weiner, Rev. Sci. Instrum. \textbf{71}, 1929 (2000).
\end{references}
\end{document}